\documentstyle[titlepage,preprint,aps,floats,epsfig]{revtex}
\newcommand{\beq}{\begin{equation}}
\newcommand{\eeq}{\end{equation}}
\newcommand{\eq}[1]{eq.(\ref{#1})}
\begin{document}
\draft
\preprint{UK/01-03}
\tighten
\title {Two-Loop Polarization Contributions to Radiative-Recoil
Corrections to Hyperfine Splitting in Muonium}
\medskip
\author {Michael I. Eides \thanks{E-mail address:
eides@pa.uky.edu, eides@thd.pnpi.spb.ru}}
\address{Department of Physics and Astronomy,
University of Kentucky,
Lexington, KY 40506, USA\\
and
Petersburg Nuclear Physics Institute,
Gatchina, St.Petersburg 188350, Russia}
\author{Howard Grotch\thanks{E-mail address: asdean@pop.uky.edu}}
\address{Department of Physics and Astronomy, University of Kentucky,
Lexington, KY 40506, USA}
\author{Valery A. Shelyuto \thanks{E-mail address:
shelyuto@vniim.ru}}
\address{D. I.  Mendeleev Institute of Metrology,
St.Petersburg 198005, Russia}
\date{July, 2001}

\maketitle

\begin{abstract}
We calculate radiative-recoil corrections of order
$\alpha^2(Z\alpha)(m/M)E_F$ to hyperfine splitting in muonium generated
by the diagrams with electron and muon polarization loops. These
corrections are enhanced by the large  logarithm of the electron-muon
mass ratio. The leading logarithm cubed and logarithm squared
contributions were obtained a long time ago. The single-logarithmic and
nonlogarithmic contributions calculated here improve the theory of
hyperfine splitting, and affect the value of the electron-muon mass
ratio extracted from the experimental data on the muonium hyperfine
splitting.
\end{abstract}

%{\it Keywords}: }

\section{Introduction. Leading Logarithmic Contributions of Order
$\alpha^2(Z\alpha)(\lowercase{m}/M)\widetilde E_F$}

\begin{figure}[ht]
\centerline{\epsfig{file=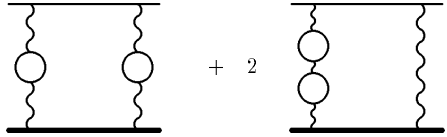,height=1.8cm}}
\vspace{0.5cm}
\caption{Graphs with two one-loop polarization insertions}
\label{onelooppolrecfhsfig}
\end{figure}

It is well known that the radiative-recoil corrections
of order $\alpha^2(Z\alpha)(m/M)\widetilde E_F$\footnote{We define the
Fermi energy  as

\beq      \label{baremuonfermi}
\widetilde{E}_{F}=\frac{16}{3}Z^4\alpha^2
\frac{m}{M} \left(\frac{m_r}{m}\right)^{3}ch\:R_{\infty},
\eeq

\noindent
where $m$ and $M$ are the electron and muon masses, $\alpha$ is the
fine structure constant, $c$ is the velocity of light, $h$ is the
Planck constant, $R_{\infty}$ is the Rydberg constant, and $Z$ is the
nucleus charge in terms of the electron charge ($Z=1$ for muonium).
The Fermi energy  $\widetilde{E}_{F}$ does not include the muon
anomalous magnetic moment $a_\mu$ which does not factorize in the case
of recoil corrections, and should be considered on the same grounds as
other corrections to hyperfine splitting.} to hyperfine splitting in
muonium are enhanced by the large logarithm of the electron-muon mass
ratio cubed \cite{es0}. The leading logarithm cube contribution is
generated by the graphs in Fig.\ \ref{onelooppolrecfhsfig}
\footnote{And by the diagrams with the crossed exchanged photon
lines. Such diagrams with the crossed exhanged photon lines are
also often omitted in other figures below.} with insertions of the
electron one-loop polarization operators in the two-photon exchange
graphs. It may be obtained almost without any calculations by
substituting the effective charge $\alpha(M)$ in the leading recoil
correction of order $(Z\alpha)(m/M)\widetilde E_F$, and expanding the
resulting expression in the power series over $\alpha$ \cite{eks89}.

\begin{figure}[ht]
\centerline{\epsfig{file=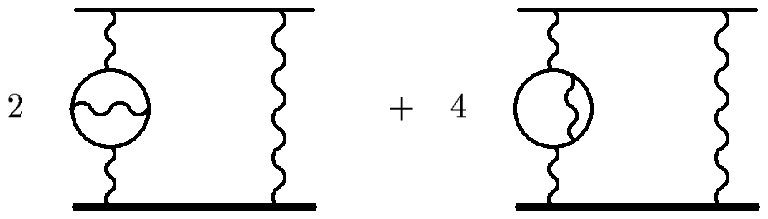,height=1.8cm}}
\vspace{0.5cm}
\caption{Graphs with two--loop polarization insertions}
\label{twolooppolrecfhsfig}
\end{figure}

\begin{figure}[ht]
\centerline{\epsfig{file=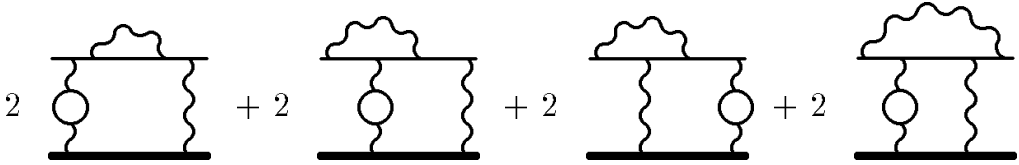,height=2.2cm}}
\vspace{0.5cm}
\caption{Graphs with radiative photon insertions}
\label{ellineinsrechfsfig}
\end{figure}

\begin{figure}[ht]
\centerline{\epsfig{file=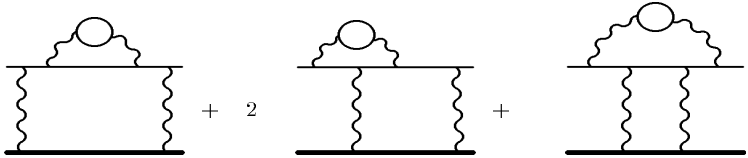,height=2.1cm,width=10cm}}
\vspace{0.5cm}
\caption{Graphs with polarization insertions in the radiative photon}
\label{polellineinsrechfsfig}
\end{figure}

\begin{figure}[ht]
\centerline{\epsfig{file=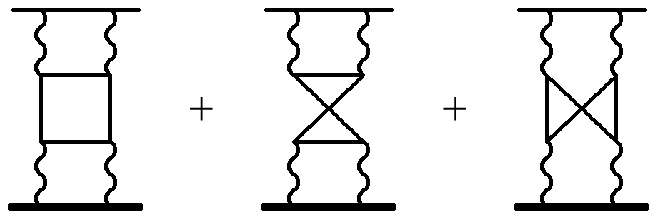,height=2.5cm}}
\vspace{0.5cm}
\caption{Graphs with light by light scattering insertions}
\label{lihlightrechfsfig}
\end{figure}

\begin{figure}[ht]
\centerline{\epsfig{file=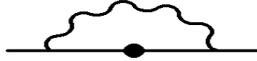,width=3.5cm, height=0.8cm}}
\vspace{0.5cm}
\caption{Renormalization of the fifth current}
\label{fifthcurhfsfig}
\end{figure}

Calculation of the logarithm squared term of order
$\alpha^2(Z\alpha)(m/M)\widetilde E_F$ is more challenging
\cite{eks89}. All graphs in Figs.\ \ref{onelooppolrecfhsfig},
\ref{twolooppolrecfhsfig}, \ref{ellineinsrechfsfig},
\ref{polellineinsrechfsfig}, and \ref{lihlightrechfsfig} generate
corrections of this order. The contribution induced by the irreducible
two-loop vacuum polarization in Fig.\ \ref{twolooppolrecfhsfig} is
again given by the effective charge expression.  Subleading logarithm
squared terms generated by the one-loop polarization insertions in
Fig.\ \ref{onelooppolrecfhsfig} may easily be calculated with the help
of the two leading asymptotic terms in the polarization operator
expansion and the skeleton integral. The logarithm squared contribution
generated by the diagrams in Fig.\ \ref{ellineinsrechfsfig} is obtained
from the leading single-logarithmic contribution of the diagrams without
polarization insertions by the effective charge substitution. An
interesting effect takes place in calculation of the logarithm squared
term generated by the polarization insertions in the radiative photon
in Fig.\ \ref{polellineinsrechfsfig}. One might expect that the high
energy asymptote of the electron factor with the polarization insertion
is given by the product of the leading constant term of the electron
factor $-5\alpha/(4\pi)$ and the leading polarization operator term.
However, this expectation turns out to be wrong.  One may check
explicitly that instead of the naive factor above one has to multiply the
polarization operator by the factor $-3\alpha/(4\pi)$. The reason for
this effect may easily be understood.  The factor $-3\alpha/(4\pi)$ is
the asymptote of the electron factor in massless QED and it gives a
contribution to the logarithmic asymptotics only after the polarization
operator insertion. This means that in massive QED the part
$-2\alpha/(4\pi)$ of the constant electron factor originates from the
integration region where the integration momentum is of order of the
electron mass. Naturally this integration region does not give any
contribution to the logarithmic asymptotics of the radiatively
corrected electron factor. The least trivial logarithm squared
contribution is generated by the three-loop diagrams in Fig.\
\ref{lihlightrechfsfig} with the insertions of the light by light
scattering block. Their contribution was calculated explicitly in
\cite{eks89}. Later it was realized that these contributions are
intimately connected with the well known anomalous renormalization of
the axial current in QED \cite{kes90}. Due to the projection on the HFS
spin structure in the logarithmic integration region the heavy particle
propagator effectively shrinks to an axial current vertex, and in this
situation calculation of the respective contribution to HFS reduces to
substitution of the well known two-loop axial renormalization factor in
Fig.\ \ref{fifthcurhfsfig} \cite{adler} in the recoil skeleton diagram.
Of course, this calculation reproduces the same contribution as
obtained by direct calculation of the diagrams with light by light
scattering expressions. From the theoretical point of view it is
interesting that one can measure anomalous two-loop renormalization of
the axial current in the atomic physics experiment.

The sum of all logarithm cubed and logarithm squared  contributions of
order $\alpha^2(Z\alpha)(m/M)\widetilde E_F$ is given by the expression
\cite{es0,eks89}

\beq
\Delta
E=\left(-\frac{4}{3}\ln^3\frac{M}{m}+\frac{4}{3}\ln^2\frac{M}{m}\right)
\frac{\alpha^2(Z\alpha)}{\pi^3}\frac{m}{M} {\widetilde E}_F.
\eeq

\noindent
It was also shown in \cite{eks89} that there are no other
contributions with the large logarithm of the mass ratio squared
accompanied by the factor $\alpha^3$, even if the factor $Z$ enters in
another manner than in the equation above.

Single-logarithmic and nonlogarithmic terms of order
$\alpha^2(Z\alpha)(m/M)\widetilde E_F$ are generated by all diagrams in
Figs.\ \ref{onelooppolrecfhsfig}-\ref{polellineinsrechfsfig}, by the
graphs with the muon polarization loops, by the graphs with
polarization and radiative photon insertions in the muon
line, and also by the graphs with two radiative photons in the electron
and/or muon lines. Only a partial result for the single-logarithmic and
nonlogarithmic corrections generated by  the pole part of the
graphs with both electron and muon polarization loops is known now
\cite{lse}. Numerically respective contribution is about 9 Hz, and may
be considered only as an indication of the scale of the respective
corrections. Corrections of this scale are phenomenologically
relevant for modern experiment and theory \cite{egs01r}. In this paper
we calculate all radiative-recoil corrections generated by the diagrams
including only the polarization loops, either electronic or muonic,
leaving calculation of the other contributions  for the future.

\section{Two-Photon Exchange Diagrams. Cancellation of the
Electron and Muon Loops}

\begin{figure}[ht]
\centerline{\epsfig{file=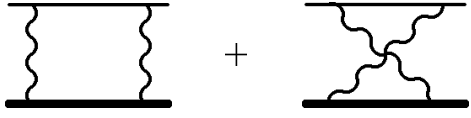,height=1.5cm}}
\vspace{0.5cm}
\caption{Diagrams with two-photon exchanges}
\label{twophothfsfig}
\end{figure}

Calculation of single-logarithmic and nonlogarithmic radiative-recoil
corrections of relative order $\alpha^2(Z\alpha)(m/M)$ (and also of
orders $(Z^2\alpha)^2(Z\alpha)(m/M)$ and
$\alpha(Z^2\alpha)(Z\alpha)(m/M)$) resembles in many respects
calculation of the corrections of relative orders
$\alpha(Z\alpha)(m/M)$ and $Z^2\alpha(Z\alpha)(m/M)$.
It was first discovered in \cite{ty,sty} that the contributions of the
diagrams with insertions of the electron and muon polarization loops
partially cancel, and, hence, it is convenient to treat such
diagrams simultaneously\footnote{We always consider the external muon
as a particle with charge $Ze$, this makes origin of different
contributions more transparent. However, somewhat inconsequently we
omit the factor $Z$ in the case of the closed muon loops. The reason
for this  apparent inconsistency is just the cancellation which we
discuss now.}. Similar cancellation holds also for the corrections of
order $\alpha^2(Z\alpha)(m/M)\widetilde E_F$, so we will first remind
the reader how it arises when one calculates the polarization
contribution of order $\alpha(Z\alpha)(m/M)\widetilde E_F$. The
nonrecoil contribution in the heavy particle pole of the two-photon
exchange diagrams exactly cancels in the sum of the electron and muon
polarizations (see for more details \cite{sty,egs01r}). Then the
skeleton recoil contribution to the hyperfine splitting generated by
the diagrams with two-photon exchanges in Fig.\ \ref{twophothfsfig} is
the result of the subtraction of the heavy pole contribution

\beq  \label{recskel}
\Delta E =4\frac{Z\alpha}{\pi}\frac{m}{M}
\widetilde{E}_F \int_{0}^{\infty}\frac{dk}{k}
\biggl[f(\mu k)-f\biggl(\frac{k}{2}\biggr)\biggr],
\eeq

\noindent
where $\mu=m/(2M)$, and

\beq
f(k)=\frac{1}{k}\biggl(\sqrt{1+k^2} - k -1\biggr)
- \frac{1}{2}\biggl( k\sqrt{1+k^2} - k^2- \frac{1}{2} \biggr),
\eeq
\[
f(k)_{k\to 0}\to -\frac{3}{4}+\frac{k^2}{2},\qquad f(k)_{k\to
\infty}\to -\frac{1}{k}.  \]

The electron polarization contribution is obtained from the skeleton
integral by multiplying the expression in \eq{recskel} by the
multiplicity factor 2, and inserting the polarization operator
$(\alpha/\pi)k^2I_1(k)$ in the integrand

\beq
\frac{\alpha}{\pi}k^2I_1(k)\equiv
\frac{\alpha}{\pi}k^2\int_0^1 {dv} \frac{v^2(1-v^2/3)}{4 + k^2(1-v^2)}.
\eeq

\noindent
The muon polarization contribution is given by a similar expression,
the only difference is that

\beq
I_1(k)\to I_{1\mu}(k)\equiv \int_0^1 {dv}
\frac{v^2(1-v^2/3)}{\mu^{-2} + k^2(1-v^2)}.
\eeq

\begin{figure}[ht]
\centerline{\epsfig{file=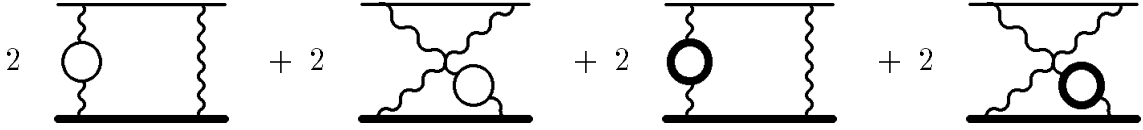,height=1.7cm}}
\vspace{0.5cm}
\caption{Diagrams with one-loop polarization insertions}
\label{photlineradrechfsfir}
\end{figure}

\noindent
Then the total recoil contribution induced by the diagrams with both
the one-loop electron and muon polarizations in Fig.\
\ref{photlineradrechfsfir} has the form

\beq \label{electrmuoncontr}
\Delta E =8\frac{\alpha(Z\alpha)}{\pi^2}\frac{m}{M}
\widetilde{E}_F \int_{0}^{\infty}\frac{dk}{k}
\biggl[f(\mu
k)-f\biggl(\frac{k}{2}\biggr)\biggr][k^2I_1(k)+k^2I_{1\mu}(k)].
\eeq

\noindent
Next we rescale the integration variable $k\to kM/m$ in the muon term
and obtain

\beq
\Delta E =8\frac{\alpha(Z\alpha)}{\pi^2}\frac{m}{M}
\widetilde{E}_F \int_{0}^{\infty}\frac{dk}{k}
\biggl[f(\mu k)-f\biggl(\frac{k}{2}\biggr)+f\biggl(\frac{
k}{2}\biggr)-f\biggl(\frac{k}{4\mu}\biggr)\biggr]k^2I_1(k)
\eeq
\[
=8\frac{\alpha(Z\alpha)}{\pi^2}\frac{m}{M}
\widetilde{E}_F \int_{0}^{\infty}\frac{dk}{k}
\biggl[f(\mu k)-f\biggl(\frac{k}{4\mu}\biggr)\biggr]k^2I_1(k).
\]

\noindent
We see that the electron and muon polarization contributions have
partially cancelled. Moreover, it is not difficult to check explicitly
that the term with $f(k/(4\mu))$ generates only corrections of higher
order in $\mu$, so with linear accuracy in the small mass ratio $m/M$
all recoil contributions generated by the diagrams with the one-loop
electron and muon polarization insertions in Fig.\
\ref{photlineradrechfsfir} are given by the integral

\beq  \label{oneloopcanc}
\Delta E =8\frac{\alpha(Z\alpha)}{\pi^2}\frac{m}{M}
\widetilde{E}_F \int_{0}^{\infty}\frac{dk}{k}f(\mu k)k^2I_1(k).
\eeq

\noindent
This integral was calculated in \cite{sty} and we will not discuss its
calculation here. Our only goal in this Section was to demonstrate the
mechanism of the partial cancellation of the electron loop and muon
loop contributions.

\section{Diagrams with either Two Electron or Two Muon Loops}

\begin{figure}[ht]
\centerline{\epsfig{file=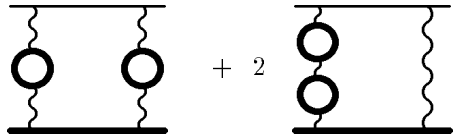,height=1.8cm,width=6cm}}
\vspace{0.5cm}
\caption{Graphs with two muon one-loop polarization insertions}
\label{eloopmuloop}
\end{figure}

The nonrecoil contribution generated by the diagrams with two electron
or muon loops in Fig.\ \ref{onelooppolrecfhsfig} and Fig.\
\ref{eloopmuloop} was obtained a long time ago \cite{eks1}. Although it
was not emphasized in that work explicitly, it is easy to check that
the result in \cite{eks1} includes heavy pole contributions which are
due to the diagrams with both the electron and muon polarizations.
Repeating the same steps as in the previous Section, it is easy to see
that the recoil contribution generated by the diagrams in Fig.\
\ref{onelooppolrecfhsfig} and Fig.\ \ref{eloopmuloop} is determined by
the integral

\beq \label{twoelmuloopan}
\Delta E =12\frac{\alpha^2(Z\alpha)}{\pi^3}\frac{m}{M}
\widetilde{E}_F \int_{0}^{\infty}\frac{dk}{k}f(\mu k)k^4I_1^2(k),
\eeq

\noindent
where the numerical factor before the integral is due to the
multiplicity of the diagrams, and the whole integral is similar to the
integral in \eq{oneloopcanc}. The only significant difference is that
now we have the two-loop factor $k^4I^2(k)$ in the integrand instead of
the one-loop factor $k^2I_1(k)$.

We calculate the integral in \eq{twoelmuloopan} separating the
contributions of small and large momenta with the help of the auxiliary
parameter $\sigma$ such that $1\ll\sigma\ll1/\mu$

\beq
\Delta E =
3(B_{11}^{<}+B_{11}^{>}) \frac{\alpha^2(Z\alpha)}{\pi^3}\frac{m}{M}
\widetilde{E}_F.
\eeq

\noindent
Then for the small integration momenta region in the leading order in
$\mu\sigma$ ($\mu k\leq\mu\sigma\ll1$) we have

\beq
B_{11}^{<}=4\int_{0}^{\sigma}\frac{dk}{k}f(\mu k)k^4I_1^2(k)
\simeq -3 \int_{0}^{\sigma}{\frac{dk}{k}}k^4I_1^2(k).
\eeq

\noindent
We substitute in this integral the closed expression for the
polarization $I_1(k)$, and again preserving only the leading
contributions in $\mu \sigma \ll 1$  obtain

\beq
B_{11}^{<}
\simeq -\frac{4}{9}\ln^3{\sigma}
+ \frac{10}{9}\ln^2{\sigma}- \frac{25}{27}\ln{\sigma}
- \frac{2}{3}\zeta{(3)}+ \frac{203}{324}.
\eeq

\noindent
The high momenta contribution is calculated by expanding the
polarization operator in $1/k^2\leq 1/\sigma^2\ll1$

\beq
B_{11}^{>}=4\int_{\sigma}^\infty\frac{dk}{k}f(\mu k)k^4I_1^2(k)
\simeq 4 \int_{\sigma}^{\infty}{\frac{dk}{k}} \biggl[
\frac{1}{\mu k}\biggl(\sqrt{1+\mu^2k^2} - \mu k - 1 \biggr)
\eeq
\[
- \frac{1}{2}\biggl(\mu k \sqrt{1+\mu^2k^2} - \mu^2 k^2
- \frac{1}{2} \biggr)\biggr]
\biggl(\frac{2}{3}\ln{k}-\frac{5}{9}\biggl)^2.
\]

\noindent
For calculation of this integral we use the standard integrals
introduced in \cite{egs98} as well as some new standard integrals (see
Appendix), and obtain

\beq
B_{11}^{>}=
\frac{4}{9}\ln^3{(2\mu)}-\frac{8}{9}\ln^2{(2\mu)}
+ \biggl(\frac{2\pi^2}{9} + \frac{25}{27} \biggr)\ln{(2\mu)}
\eeq
\[
+\frac{2}{3}\zeta{(3)}-\frac{4\pi^2}{27}- \frac{41}{18}
+\frac{4}{9}\ln^3{\sigma}-\frac{10}{9}\ln^2{\sigma}
+ \frac{25}{27}\ln{\sigma}.
\]

\noindent
Now we are ready to write down the total recoil contribution generated
by the diagrams in Fig.\ \ref{onelooppolrecfhsfig} and Fig.\
\ref{eloopmuloop}

\beq \label{finemuoneloop}
\Delta E
=\biggl[-\frac{4}{3}\ln^3{\frac{M}{m}}-\frac{8}{3}\ln^2{\frac{M}{m}}
- \biggl(\frac{2\pi^2}{3} + \frac{25}{9} \biggr)\ln{\frac{M}{m}}
- \frac{4\pi^2}{9} - \frac{535}{108}\biggr]
\frac{\alpha^2(Z\alpha)}{\pi^3}\frac{m}{M}\widetilde{E}_F.
\eeq

\noindent
The logarithm cube and logarithm squared terms in this expression are
already known \cite{es0,eks89}, and the single-logarithmic and
nonlogarithmic terms are obtained here.

\section{Diagrams with both the Electron and Muon
Loops}\label{mixedidgrsect}

\begin{figure}[ht]
\centerline{\epsfig{file=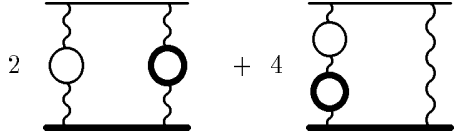,height=1.8cm,width=6cm}}
\vspace{0.5cm}
\caption{Graphs with both the electron and muon loops}
\label{mixedloops}
\end{figure}

Consider now the diagrams with one electron and one muon loop in Fig.\
\ref{mixedloops}. We can look at these diagrams as a result of the
electron polarization operator insertions in the muon loop diagrams in
Fig.\ \ref{photlineradrechfsfir}. The complete analytic expression for
the last two diagrams in Fig.\ \ref{photlineradrechfsfir} has the
form

\beq  \label{totalmuonone}
\Delta E =8\frac{\alpha(Z\alpha)}{\pi^2}\frac{m}{M}
\widetilde{E}_F \int_{0}^{\infty}\frac{dk}{k}
\biggl[\widetilde f(\mu k)-\widetilde f\biggl(\frac{k}{2}\biggr)
\biggr]k^2I_{1\mu}(k),
\eeq

\noindent
where

\beq
\widetilde f(k)=f(k)+\frac{1}{k}.
\eeq

\noindent
Unlike \eq{electrmuoncontr} we have restored in \eq{totalmuonone} the
heavy particle pole contribution, which in the case of the muon
polarization loop also carries the recoil factor. To simplify further
calculations we rescale the integration momentum $k\to kM/m$

\beq
\Delta E =8\frac{\alpha(Z\alpha)}{\pi^2}\frac{m}{M}
\widetilde{E}_F \int_{0}^{\infty}\frac{dk}{k}
\biggl[\widetilde f\biggl(\frac{k}{2}\biggr)-\widetilde
f\biggl(\frac{k}{4\mu}\biggr) \biggr]k^2I_1(k),
\eeq

\noindent
and note that with the linear accuracy in $m/M$ we may omit the second
term in the square brackets in the integrand. Then the muon loop
diagrams in Fig.\ \ref{photlineradrechfsfir} are described by the
expression

\beq  \label{1muloop}
\Delta E =8\frac{\alpha(Z\alpha)}{\pi^2}\frac{m}{M}
\widetilde{E}_F \int_{0}^{\infty}\frac{dk}{k}
\widetilde f\biggl(\frac{k}{2}\biggr)k^2I_1(k).
\eeq

\noindent
The integral in \eq{1muloop} turns into the contribution of the
diagrams in Fig.\ \ref{mixedloops} after multiplication by the factor 3
and insertion in the integrand of the additional factor

\beq
\frac{\alpha}{\pi}\biggl(\frac{k}{2\mu}\biggr)^2I_1(\frac{k}{2\mu})
=\frac{\alpha}{\pi}
\biggl[\frac{2}{3}\ln\frac{k}{2\mu}-\frac{5}{9}+O(\frac{\mu^2}{k^2})\biggr].
\eeq

\noindent
This extra factor enters in the asymptotic regime
since the characteristic scale of the integration momenta in
\eq{1muloop} is about one, and the parameter $\mu$ goes to zero.

Then the contribution to HFS of the diagrams in Fig.\ \ref{mixedloops}
is given by the integral

\beq
\Delta E=24\frac{\alpha^2(Z\alpha)}{\pi^3}\frac{m}{M}
\widetilde{E}_F \int_{0}^{\infty}\frac{dk}{k}
\widetilde
f\biggl(\frac{k}{2}\biggr)k^2I_1(k)\biggl(\frac{2}{3}\ln\frac{k}{2\mu}
-\frac{5}{9}\biggr)
\eeq
\[
=
24 \frac{\alpha^2(Z\alpha)}{\pi^3}
\frac{m}{M}\widetilde{E}_F\int_{0}^{\infty}{dk} \biggl[
\biggl(\sqrt{4+k^2} - k \biggr)
\]
\[
- \frac{k}{2}\biggl(\frac{k}{4} \sqrt{4+k^2} - \frac{k^2}{4}
- \frac{1}{2} \biggr)
\biggr]\int_0^1 {dv} \frac{v^2(1-v^2/3)}{4+k^2(1-v^2)}
\biggl[\frac{2}{3} \ln{\frac{M}{m}}
+ \frac{2}{3} \ln{k}- \frac{5}{9}  \biggr].
\]

\noindent
After calculation we obtain (see Appendix)

\beq   \label{finmixoneloop}
\Delta E=
\biggl[
\biggl(\frac{2\pi^2}{3} - \frac{20}{9} \biggr) \ln{\frac{M}{m}}
+ \frac{\pi^2}{3} - \frac{53}{9} \biggr]\frac{\alpha^2(Z\alpha)}{\pi^3}
\frac{m}{M}\widetilde{E}_F.
\eeq

\section{Diagrams with Second Order Polarization
Insertions}\label{twolooppolsect}

\begin{figure}[ht]
\centerline{\epsfig{file=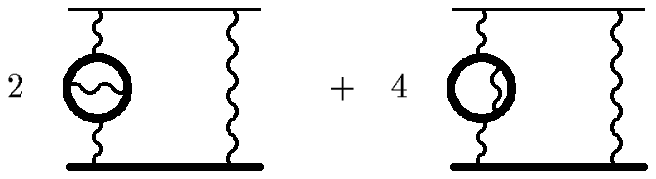,height=1.8cm}}
\vspace{0.5cm}
\caption{Graphs with muon two-loop polarization insertions}
\label{twoelmuloopanfig}
\end{figure}

The recoil contribution to HFS generated by the diagrams in
Fig.\ \ref{twolooppolrecfhsfig} and Fig.\ \ref{twoelmuloopanfig} with
two-loop electron and muon polarization insertions is given by the
integral (compare \eq{oneloopcanc})

\beq  \label{twolooppolan}
\Delta E=
8\frac{\alpha^2(Z\alpha)}{\pi^3}\frac{m}{M}
\widetilde{E}_F \int_{0}^{\infty}\frac{dk}{k}f(\mu k)k^2I_2(k),
\eeq

\noindent
where $(\alpha^2/\pi^2)k^2I_2(k)$ is the two-loop polarization operator
\cite{kalsab,schwinger}

\beq
I_2(k) = \frac{2}{3} \int_0^1 {dv}\frac{v}{4+k^2(1-v^2)}
\biggl\{(3-v^2)(1+v^2)\biggl[\mbox{Li}_2 \biggl( -\frac{1-v}{1+v}\biggr)
\eeq
\[
+ 2\mbox{Li}_2 \biggl(\frac{1-v}{1+v}\biggr)
+\frac {3}{2} \ln{\frac{1+v}{1-v}} \ln{\frac{1+v}{2}}
- \ln{\frac{1+v}{1-v}} \ln{v} \biggr]
\]
\[
+ \biggl[\frac{11}{16}(3-v^2)(1+v^2) + \frac{v^4}{4}\biggr]
\ln{\frac{1+v}{1-v}}
\]
\[
+\biggl[\frac{3}{2}v(3-v^2)\ln{\frac{1-v^2}{4}}
- 2v(3-v^2)\ln{v} \biggr] + \frac {3}{8}v(5-3v^2)\biggl\}.
\]

To simplify further calculations we represent the two-loop polarization
operator in the form

\beq
\label{trick}
I_2(k)= \frac{3}{4}I_1(k)+ \int_0^1 {dv}
\frac{R(v)}{4+k^2(1-v^2)},
\eeq

\noindent
where

\beq  \label{rvpol}
R(v)= \frac{2}{3} v \biggl\{(3-v^2)(1+v^2)\biggl[\mbox{Li}_2 \biggl(
-\frac{1-v}{1+v}\biggr)
\eeq
\[
+ 2\mbox{Li}_2 \biggl(\frac{1-v}{1+v}\biggr)
+\frac {3}{2} \ln{\frac{1+v}{1-v}} \ln{\frac{1+v}{2}}
- \ln{\frac{1+v}{1-v}} \ln{v} \biggr]
\]
\[
+ \biggl[\frac{11}{16}(3-v^2)(1+v^2) + \frac{v^4}{4}\biggr]
\ln{\frac{1+v}{1-v}}
\]
\[
+\biggl[\frac{3}{2}v(3-v^2)\ln{\frac{1-v^2}{4}}
- 2v(3-v^2)\ln{v} \biggr]+ \frac{3}{4} v(1-v^2)\biggl\}.
\]

\noindent
The integral in \eq{trick} decreases as $1/k^2$ at large $k$.
Note that since $R(v)\to 3(1-v)$ at $v \to 1$ this leading term in
the asymptotic expansion is not enhanced by the large logarithm $\ln
k$.  Absence of this logarithm significantly simplifies further
calculations.

In terms of the function $R(v)$ the integral for the recoil
contribution in \eq{twolooppolan} has the form

\beq
\Delta E=
8\frac{\alpha^2(Z\alpha)}{\pi^3}\frac{m}{M}
\widetilde{E}_F \int_{0}^{\infty}{dk}{k}f(\mu k)
\biggl[\frac{3}{4}I_1(k)+ \int_0^1 {dv}
\frac{ R(v)}{4+k^2(1-v^2)}\biggr]
\eeq
\[
\equiv \Delta E^{a} + \Delta E^{b}.
\]

\noindent
The first contribution on the right hand side is proportional to
the well known one-loop contribution in \eq{oneloopcanc} \cite{cp,ty,sty}

\beq      \label{Ea}
\Delta E^{a}=\biggl[-\frac{3}{2}\ln^2{(2\mu)} + 2\ln{(2\mu)}
- \frac{\pi^2}{4} - \frac{7}{3} \biggr]
\frac{\alpha^2(Z\alpha)}{\pi^3}\frac{m}{M}\widetilde{E}_F.
\eeq

Calculation of the second term $\Delta E^{b}$ is a bit more involved.
We again introduce the auxiliary parameter $\sigma$
($1\ll\sigma\ll1/\mu$) and consider separately the small and large
momenta contributions. For the small integration momenta region
in the leading order in $\mu\sigma$ we have

\beq  \label{eb<}
\Delta E^{b<}=8\frac{\alpha^2(Z\alpha)}{\pi^3}\frac{m}{M}
\widetilde{E}_F \int_{0}^{\sigma}{dk}{k}f(\mu k)
\int_0^1 {\frac{dv R(v)}{4+k^2(1-v^2)}}
\eeq
\[
\simeq -3\frac{\alpha^2(Z\alpha)}{\pi^3}\frac{m}{M}\widetilde{E}_F
\int_{0}^{\sigma}{dk^2}\int_0^1 {\frac{dv R(v)}{4+k^2(1-v^2)}}
\simeq-3\frac{\alpha^2(Z\alpha)}{\pi^3}\frac{m}{M}E_F
\int_0^1 {dv} \frac{R(v)}{1-v^2}
\ln{\frac{\sigma^2(1-v^2)}{4}}
\]
\[
=-3\frac{\alpha^2(Z\alpha)}{\pi^3}\frac{m}{M}\widetilde{E}_F
\Biggl\{\biggl[\zeta (3) + \frac{5}{24}
\biggr] \ln{\frac{\sigma^2}{4}}+
2\zeta{(3)}\ln{2}
+\frac{25}{24}\zeta{(3)} +
\frac{16}{3}\mbox{Li}_4\biggl(\frac{1}{2}\biggr)
\]
\[
-\frac{2\pi^2}{9}\ln^2{2} + \frac{2}{9}\ln^4{2}+\frac{5}{12}\ln{2}
- \frac{5\pi^4}{108}- \frac{223}{144}
\Biggr\},
\]

\noindent
where we used certain integrals for the function $R(v)$ collected in
the Appendix.

The high-momentum contribution is given by the integral

\beq
\Delta E^{b>}=8\frac{\alpha^2(Z\alpha)}{\pi^3}\frac{m}{M}
\widetilde{E}_F \int_{\sigma}^\infty{dk}{k}f(\mu k)
\int_0^1 {\frac{dv R(v)}{4+k^2(1-v^2)}}.
\eeq

\noindent
First we use that $k^2\gg\sigma^2\gg1$ and that $R(v)\to3(1-v)$ as
$v\to1$ and omit $4$ in the denominator in the integrand, and
then perform the calculations using the integrals
from the Appendix

\beq             \label{eb>}
\Delta E^{b>}=4\biggl[\zeta (3) + \frac{5}{24}\biggr]
\frac{\alpha^2(Z\alpha)}{\pi^3}\frac{m}{M}\widetilde{E}_F
\int_{\sigma^2}^{\infty}{\frac{dk^2}{k^2}}\biggl[
\frac{1}{\mu k}\biggl(\sqrt{1+\mu^2k^2} - \mu k - 1 \biggr)
\eeq
\[
- \frac{1}{2}\biggl(\mu k \sqrt{1+\mu^2k^2} - \mu^2 k^2
- \frac{1}{2} \biggr)\biggr]
\]
\[
=2 \biggl[\zeta (3) + \frac{5}{24}\biggr]
\biggl[3\ln{(2\mu)} -\frac{9}{2}+ 3\ln{\sigma}\biggl]
\frac{\alpha^2(Z\alpha)}{\pi^3}\frac{m}{M}\widetilde{E}_F.
\]

The total recoil contribution to HFS generated by the diagrams with
two-loop polarization insertions in Fig.\ \ref{twoelmuloopanfig} is
given by the sum of the contributions in \eq{Ea}, \eq{eb<}, and \eq{eb>}

\beq   \label{fintwoloop}
\Delta E=\Delta E^{a}+\Delta E^{b<}+ \Delta E^{b>}
\eeq
\[
= \biggl\{
-\frac{3}{2} \ln^2{\frac{M}{m}}
- \biggl[6\zeta (3) + \frac{13}{4}\biggr]\ln{\frac{M}{m}}
-\frac{97}{8}\zeta{(3)} - 16\mbox{Li}_4\biggl(\frac{1}{2}\biggr)
\]
\[
+ \frac{2\pi^2}{3}\ln^2{2} - \frac{2}{3}\ln^4{2}
+ \frac{5\pi^4}{36} - \frac{\pi^2}{4} + \frac{7}{16}\biggr\}
\frac{\alpha^2(Z\alpha)}{\pi^3}
\frac{m}{M}\widetilde{E}_F.
\]

\noindent
The logarithm squared term in this expression was obtained
in \cite{eks89}, and the single-logarithmic and nonlogarithmic terms are
obtained here.

\section{Discussion of Results}

Collecting all contributions in \eq{finemuoneloop}, \eq{finmixoneloop},
and \eq{fintwoloop} we obtain

\beq  \label{totalrecpol}
\Delta E_t=
\biggl\{-\frac{4}{3}\ln^3\frac{M}{m}-\frac{25}{6}\ln^2\frac{M}{m}
-\biggl[6\zeta(3)+\frac{33}{4}\biggr]\ln\frac{M}{m}
\eeq
\[
- \frac{97}{8} \zeta(3)- 16\mbox{Li}_4\biggl(\frac{1}{2}\biggr)
+\frac{2\pi^2}{3}\ln^22-\frac{2}{3} \ln^42+ \frac{5\pi^4}{36}
- \frac{13\pi^2}{36}-\frac{4495}{432}\biggr\}
\frac{\alpha^2(Z\alpha)}{\pi^3}\frac{m}{M}\widetilde{E}_F.
\]

\noindent
The contribution which contains only single logarithms and constants is

\beq     \label{result}
\Delta E=
\biggl\{-\biggl[6\zeta(3)+\frac{33}{4}\biggr]\ln\frac{M}{m}
\eeq
\[
- \frac{97}{8} \zeta(3)- 16\mbox{Li}_4\biggl(\frac{1}{2}\biggr)
+\frac{2\pi^2}{3}\ln^22-\frac{2}{3} \ln^42+ \frac{5\pi^4}{36}
- \frac{13\pi^2}{36}-\frac{4495}{432}\biggr\}
\frac{\alpha^2(Z\alpha)}{\pi^3}\frac{m}{M}
\widetilde{E}_F.
\]

The factor in the square brackets is about $(-103)$, and numerically
the respective contribution to the muonium HFS is

\beq     \label{number}
\Delta E_{new}=-0.027~7~\mbox{kHz}.
\eeq

The magnitude of this correction is just in the range we should
expect based on the partial result in \cite{lse}. The contribution in
\eq{number} is of the same scale as the logarithmic in
$Z\alpha$ corrections of order $(Z\alpha)^3(m/M)E_F$ and
$\alpha(Z\alpha)^2(m/M)E_F$, calculated recently in \cite{my,rh}.

Collecting the recent results from \cite{egs98,my,rh} and \eq{result},
and using the experimental value of the muonium hyperfine splitting
\cite{lbdd} we may derive a value of the electron-muon mass ratio

\beq  \label{masssepar}
\frac{M}{m}=206.768~279~8~(23)~(16)~(32),
\eeq

\noindent
where the first error comes from the experimental error of the
hyperfine splitting measurement, the second comes from the error in the
value of the fine structure constant $\alpha$, and the third is an
estimate of the yet unknown theoretical contributions.

Combining all errors we  obtain the mass ratio

\beq    \label{masspred}
\frac{M}{m}=206.768~279~8~(43),\qquad \delta=2\cdot 10^{-8},
\eeq

\noindent
which is almost six times more accurate than the best direct
experimental value in \cite{lbdd}.

Estimating the errors in \eq{masssepar} we assumed that the
theoretical error of calculation of the muonium hyperfine splitting is
about 70 Hz. This theoretical error is determined by the estimate of
the still uncalculated terms which include single-logarithmic and
nonlogarithmic radiative-recoil corrections  of order
$\alpha^2(Z\alpha)(m/M)\widetilde E_F$ generated by the graphs containing
besides the polarization loops also radiative photons, as well as the
nonlogarithmic contributions of order $(Z\alpha)^3(m/M)E_F$,
$\alpha(Z\alpha)^2(m/M)E_F$, and  some other corrections (see a more
detailed analysis in \cite{egs01r,mt00}).  Calculation of all these
contributions and reduction of the theoretical uncertainty of the
hyperfine splitting in muonium below 10 Hz is the current task of the
theory. As the next step towards this goal we hope to present soon the
results of calculation of the single-logarithmic and nonlogarithmic
radiative-recoil corrections of order $\alpha^2(Z\alpha)(m/M)\widetilde
E_F$ generated by the graphs containing besides the polarization loops
also radiative photons.

\acknowledgements

This work was supported by the NSF grant PHY-0049059. Work of V. A. S.
was also supported in part by the RFBR grant \# 00-02-16718.

\appendix
\section{Auxiliary Integrals}

All contributions to hyperfine splitting in the main body of this paper
are written with the help of the function

\beq
f(\mu k) \equiv \Phi_0^{\mu} (k)  + \frac{1}{2} \Phi_1^{\mu} (k),
\eeq

\noindent
where the standard auxiliary functions $\Phi_n (k)$ where introduced in
\cite{egs98})

\begin{equation}
\Phi_0^{\mu} (k)  = W(\xi_{\mu}) - \frac{1}{\sqrt{\xi_{\mu}}},
\end{equation}
\begin{equation}
\Phi_1^{\mu} (k)  = -\xi_{\mu}  W(\xi_{\mu}) + \frac{1}{2},
\end{equation}

\noindent
and
\begin{equation}
W(\xi_{\mu}) = \sqrt{1+\frac{1}{\xi_{\mu}}} - 1,\qquad\xi_{\mu} =
\mu^2k^2.
\end{equation}

\noindent

\noindent
In terms of these functions all high-momentum contributions to
hyperfine splitting may be represented as linear combinations of the
standard integrals

\begin{equation}
V_{lmn} = \int_{\sigma}^{\infty} \frac{d k^2}{(k^2)^l}
 (\ln k )^m  \Phi_n^{\mu} (k),
\end{equation}

\noindent
where $l=1$, $m=0,1,2$ and $n=0,1$. Calculation of these integrals was
described in \cite{egs98}, and we present here only the results
for the two integrals which were not calculated in \cite{egs98}

\beq
V_{120}= \frac{2}{3}  \ln^3{(2\mu)}- 2\ln^2{(2\mu)}
+ \left(\frac{\pi^2}{3} + 4 \right)  \ln{(2\mu)}
+ \zeta{(3)} - \frac{\pi^2}{3}- 4
+ \frac{2}{3}  \ln^3{\sigma},
\eeq
\beq
V_{121} = - \frac{1}{3} \ln^3{(2\mu)}
- \frac{1}{2} \ln^2{(2\mu)}
- \left(\frac{\pi^2}{6} + \frac{1}{2} \right) \ln{(2\mu)}
- \frac{1}{2} \zeta{(3)}- \frac{\pi^2}{12} - \frac{1}{4}
- \frac{1}{3} \ln^3{\sigma}.
\eeq

In Section \ref{mixedidgrsect} we encountered the integral

\beq
\int_{0}^{\infty}{dk} \ln{k} \biggl[
\biggl(\sqrt{4+k^2} - k \biggr)
- \frac{k}{8}\biggl(k \sqrt{4+k^2} - k^2
- 2 \biggr)
\biggr]\int_0^1 {dv} \frac{v^2(1-v^2/3)}{4+k^2(1-v^2)}
\eeq
\[
=\frac{\pi^2}{18} - \frac{209}{432},
\]

\noindent
which may be calculated by changing the integration variable

\beq
z = \frac{2}{k+\sqrt{4+k^2}}.
\eeq

A number of integrals with the function $R(v)$ (see \eq{rvpol}) used in
Section \ref{twolooppolsect} are collected below

\beq
\int_0^1 {dv} \Bigl[\frac{3}{4} v^2 \Bigl(1 - \frac{v^2}{3}\Bigr)
+ R(v)\Bigr] = \frac{82}{81},
\eeq
\beq
\int_0^1 {dv} R(v) =\frac{329}{405},
\eeq
\beq
\int_0^1 {dv} \frac{R(v)}{1-v^2} = \zeta{(3)} + \frac{5}{24},
\eeq
\beq
\int_0^1 {dv} \frac{R(v)}{1-v^2}\ln{(1-v^2)} = 2\zeta{(3)} \ln{2}
+ \frac{25}{24}\zeta{(3)} +
\frac{16}{3}\mbox{Li}_4\biggl(\frac{1}{2}\biggr)
\eeq
\[
-
\frac{2\pi^2}{9}\ln^2{2}+ \frac{2}{9}\ln^4{2}+\frac{5}{12}\ln{2} -
\frac{5\pi^4}{108}- \frac{223}{144}.
\]

\end{document}